\documentclass[a4paper,10pt]{article}
\usepackage[a4paper, lmargin=1.0in, rmargin=1.0in, tmargin=1.0in, bmargin=1.0in]{geometry}

\usepackage{graphicx}
\usepackage{amsmath,amssymb}

\def \ep {{\epsilon}}

\newcommand{\mbR}{{\mathbb R}}
\newcommand{\mbN}{{\mathbb N}}

\newcommand{\comp}[1]{\ensuremath{[\!\!| #1 |\!\!]}}
\newcommand{\dig}[1]{\ensuremath{[ #1 ]}}

\newtheorem{Fact}{Fact}

\title{A Note on Digitization of Real-Time Models and Logics}
\author{Janardan Misra \\ janardanmisra@acm.org}
\date{}

\begin{document}
\maketitle

\subsection*{Introduction}


A real-time system behavior is commonly represented using {discrete trace} formalism in terms of an enumerable sequence of snapshots of the system states together with a measurement of the time points at which these snapshots were taken. There are two common semantic choices for modeling the time associated with the states. The first choice involves using a discrete domain and is known as {\it discrete semantics} or {\it integral semantics} (when time is $\mbN$) and second choice is when time is modeled using a dense domain e.g., using $\mbR$ and is known as {\it dense semantics}.   

Dense time semantics is strictly more expressive than discrete time and there are systems for which dense semantics is naturally preferred over discrete semantics~\cite{alur1994theory}. Examples include asynchronous circuits or systems which are composed of purely asynchronous timed components. However, many of the decision problems related to real-time system verification under dense time semantics are known to be undecidable~\cite{AH93,alur1996benefits}. On the other hand, even though, the alternative discrete semantics does not correspond very well to the actual ``physical'' behavior of the real-time systems, most of these decision problems, which were undecidable under dense semantics, become decidable and often tractable. Therefore it is highly desirable to have a technique which enables verifying properties of real-time systems under dense semantics but uses the procedures developed for discrete semantics. 

Digitization offers an effective solution to this. Digitization provides a sound and complete method to reduce the problem of verifying whether a real-time system satisfies a property under dense time semantics to whether the same real-time system satisfies the property over discrete time. Digitization was first introduced by Henzinger et al. in~\cite{HMP92} and has been further studied and applied in practical scenarios by many other authors~\cite{ouaknine2003revisiting,el2003timed,bozga1999efficient} owing to the simplification and scalability it offers for real-time system verification. 

\subsection*{Defining Digitization}


We shall adopt the terminology from~\cite{HMP92}. 
Let us define an {\em observation} to be a pair $(\sigma_i, T_i)$, where $\sigma_i$ is a state and $T_i \in T$ is the time point when $\sigma_i$ is observed. Let $Time$ be the model of time such that $T \subseteq Time$. A {\em timed state sequence} $\eta = (\sigma, T)$ is an infinite sequence $\eta: \, \, (\sigma_0, T_0) \rightarrow (\sigma_1, T_1) \rightarrow (\sigma_2, T_2) \rightarrow \cdots$ of observations. Further, the infinite sequence $T_i \in T$ of time stamps in $\eta$ satisfy (i) {\em (weak) monotonicity}: $T_i \leq T_{i+1}$ for all $i \geq 0$, and (ii) {\em progress or non-zenoness:} time progresses, for all $\delta \in Time$, $T_i \geq \delta$ for some $i \geq 0$. Note that the weak monotonicity allows multiple events to occur simultaneously at the same time point.  

A timed state sequence under dense-time model will be referred to as {\em precisely timed} and under integral-time model as  {\em digitally timed}.

Every real-time system model $S$ defines a real-time property, denoted as $\comp{S}$, which is the set of all timed state sequences of $S$. Also, every real-time specification $\phi$ defines a real-time property $\comp{\phi}$, the set of real-time sequences that satisfy $\phi$.

A real-time system $S$ is said to satisfy the specification $\phi$, written as $$ S \models \phi \mbox{ iff } {\comp{S}} \subseteq {\comp{\phi}}$$ 



Given $x \in \mbR$ and $\ep \in (0,1]$, define $\dig{x}_{\ep} = \lfloor x \rfloor$ if $x \leq \lfloor x \rfloor + \ep$, otherwise $\dig{x}_{\ep} =\lceil x \rceil$\footnote{where
$\lfloor{\cdot}\rfloor$ and $\lceil{\cdot}\rceil$ are the floor and ceiling rounding operations on real numbers respectively}. Given a precisely timed sequence $\eta = (\sigma, T)$ and $\ep \in (0,1]$, define $\ep$-{\em digitization} $\dig{\eta}_{\ep} = (\sigma, \dig{T}_{\ep})$ of $\eta$ be the digitally timed sequence \[ (\sigma_0, \dig{T_0}_{\ep}) \rightarrow (\sigma_1, \dig{T_1}_{\ep}) \rightarrow \cdots, \] 
For any dense-time property $\Pi$ (a set of timed sequences over dense time) let $$\dig{\Pi} = \{\dig{\eta}_{\ep} \,|\, \eta \in \Pi \,\, \mbox{and} \,\, \ep \in (0,1] \},$$ which
is a digitization of $\Pi$. We write $\dig{\eta}$ instead of $\dig{\{\eta\}}$.

Let $\Pi$ be a dense-time property. $\Pi$ is {\em closed under digitization} iff for all $\eta \in \Pi$ implies $\dig{\eta} \subseteq \Pi$. $\Pi$ is {\em closed under inverse digitization} iff $\dig{\eta} \subseteq \Pi$ implies $\eta \in \Pi$, for all precisely time state sequences $\eta$. Finally, $\Pi$ is {\em digitizable} iff it is closed under both digitization and inverse digitization, {\it i.e.,} $\eta \in \Pi$ iff $\dig{\eta} \subseteq \Pi$ for all precisely timed state sequences. Following key result was proved in~\cite{HMP92}:
\begin{Fact}
Assume a real-time system $S$ whose dense-time semantics $\comp{S}_{\mbR}$ is closed under digitization, and a specification $\phi$ whose dense-time semantics $\comp{\phi}_{\mbR}$ is closed under inverse digitization. Then in order to prove $S \models_{\mbR} \phi$ it suffices to check if $S \models_{\mbN} \phi$.
\end{Fact}
A dense-time property $\Pi$ is said to be {\em qualitative} if $\eta \in \Pi$ implies $\eta' \in \Pi$ for all precisely timed sequences $\eta$ and $\eta'$ with identical state components. 

\subsection*{Digitization results for Timed Automata and Real-time Temporal Logics}

Let us consider the case when timed automata (TA) will be used an implementation language. In~\cite{EOA98,Bos99,ouaknine2003revisiting} digitization of TA was considered. Let a Closed TA be the one in which all clock regions (constraints) are closed. Syntactically, a Closed TA would only involve (atomic) clock constraints of the form $x \leq c | x \geq c$. Following is an important positive result known for the digitization of Closed TA, which in turn enables verifying their dense time properties under integer time domain. 

\begin{Fact} Under dense-time (trace) semantics, Closed TA is closed under digitization. \end{Fact}

In contrast to a Closed TA, an Open TA is one which only involves clock constraints of the form $x < c | x> c$. There are further digitization related results known for both Closed as well as Open TAs:

\begin{Fact} Under dense-time (trace) semantics, Open TA is closed under inverse digitization. \end{Fact}

\begin{Fact} Checking whether a TA is closed under digitization is decidable under dense-time (trace) semantics. \end{Fact}

\begin{Fact} Checking whether a TA is closed under inverse digitization is undecidable under dense-time (trace) semantics. \end{Fact}

Apart from the trace semantics, digitization of TA has also been studied under {\it robust semantics}. Informally~\cite{alur2004decision}, under robust semantics, a TA accepts a timed sequence if and only if it also accepts the ``dense'' subset around the sequence. Following results are known for digitization under robust semantics:

\begin{Fact} The problem of closure under digitization is undecidable for TA under the robust semantics. \end{Fact} 

\begin{Fact} Under the robust semantics TA are closed under inverse digitization. \end{Fact} 

In the light of these results, for practical purposes, we can only use Closed TA as a framework for representing real-time system designs and for specifying desired properties of these models, we need to use other formalisms e.g., real-time temporal logics. A key digitization result valid for any real-time logic with discrete trace semantics was proved in~\cite{HMP92}: 

\begin{Fact} Under dense-time discrete trace semantics, all qualitative properties are closed under both digitization as well as inverse digitization. \end{Fact}

Using these facts, it follows that if we consider a closed TA $S$ and a qualitative real-time property $\phi$, it is sufficient to consider only the integral semantics for $S$ as well as $\phi$ to check whether $\phi$ is satisfied over $S$ under dense semantics.
  
For Metric Temporal Logic (MTL), following results are proved in~\cite{HMP92}:

\begin{Fact} Bounded invariance and bounded response properties are closed under digitization as well as inverse digitization. \end{Fact}
\begin{Fact}\label{f} Under dense-time (trace) semantics weakly constrained MTL formulas are closed under inverse digitization. \end{Fact} 

Where an MTL formula $\psi$ is termed {\it weakly constrained} if and only if it satisfies the following conditions: i) Every negation in $\psi$ is in front of an atomic proposition, ii) every $until$ operator in $\psi$ is constrained by an open interval, and iii) every $unless$ operator in $\psi$ is constrained by an closed interval. 

The fact~\ref{f} is used in~\cite{HMP92} to define a conservative approximation scheme for reasoning about arbitrary dense time MTL properties using digital clock models, when timed transition systems are considered as the implementation language.

Apart from the case of MTL, few other logics have also been investigated for their digitization properties. \cite{ouaknine2002digitisation,el2003timed} consider the case of {\it timed-CSP} and prove that all timed CSP formulas are closed under inverse digitization. This enabled devising a model checking algorithm (including for liveness properties) using the untimed CSP model checker FDR~\cite{Fischer99}. Similarly~\cite{chakravorty2003digitizing} have studied interval duration calculus, although, using a different notion of {\it strong closure under inverse digitization} entailing the standard notion of closure under inverse digitization.  

\subsection*{Value of Digitization}


Many instances of the verification problem for TA under dense semantics are undecidable~\cite{alur2004decision}. For example,

\begin{Fact} Under dense-time semantics, universality, equivalence, and inclusion problem for TA are undecidable. \end{Fact}

Some of these become decidable under integer semantics primarily because of the following result~\cite{alur2004decision}:

\begin{Fact} Under integral time semantics, the set of timed traces accepted by a TA is regular. \end{Fact}

In~\cite{beyer2001improvements}, it is proved that for closed TA, integer semantics is equivalent to dense semantics with respect to all reachable locations.

For the case of real-time logics, dense-time semantics introduces additional difficulties, whereas with discrete time interpretation, results and techniques for classic untimed logics can often be adapted to the real-time case~\cite{AH93}. As pointed out in~\cite{Furia07}, {\it expressive robustness} is yet another aspect where discrete time semantics offers advantages over dense time interpretation for real-time logics.    


Apart from the theoretical significance mentioned above, digitization offers distinct advantages in practice. As compared to the tools available for dense time modeling and verification of real-time systems, there exist significantly more efficient off-the-shelf tools and techniques for discrete time verification. This situation renders a dense-time verification process based on digitization very appealing from a practical viewpoint, because it can be implemented easily and it can rely on solid and scalable methods.

Digitization can significantly reduce the overall cost of verification since it allows working with a simpler time domain preserving the full verification equivalence. For example, a TA with integral semantics can be analyzed not only using the data structures suggested for dense time (e.g., DBM) but also using enumerative and symbolic techniques (such as BDDs~\cite{beyer2003can}) which cannot be applied directly under dense time semantics. The tool Rabbit~\cite{beyer2003rabbit}, for example, uses digitization in order to render BDD based symbolic techniques amenable for real-time system verification. 

Digitization based experiments with actual realistic examples~\cite{EOA98,bozga1999efficient}, applications~\cite{larsen2005online,blom2008simulated}, and tools~\cite{eshuis2004tool} clearly demonstrate that verification of real-time systems using digitization is not only feasible but scales much better as compared to using models and tools with dense time semantics.

\subsection*{Limitations of Digitization}

Existing digitization results are based upon discrete trace semantics for both implementation as well as specification languages. However, under trace semantics, branching time behaviors cannot be captured directly~\cite{HMP92}. Also when we consider TA as an implementation language, restriction of Closed TA might be bit obtrusive. However, as suggested in~\cite{ouaknine2003revisiting}, such restriction is not very limiting in practice because any open clock constraint of the form $x < c|x> c$ can always be safely infinitesimally over-approximated by $x \leq c|x \geq c$. Such over approximation renders any TA to be closed under digitization. However, the price to be paid by such over approximation, is the possible presence of false negatives during emptiness testing~\cite{Bos99}. 

\subsection*{Alternative Techniques}

\textsf{Sampling Invariance}~\cite{furia2008automated,furia2010theory} and \textsf{GoAbstraction}~\cite{clarke2007abstraction} are two important alternatives which aim to reduce dense time verification problem to discrete time verification problem.
  
%
%
%
%
%

\bibliographystyle{plain}
\bibliography{RP}

\end{document}